\documentclass[conference]{IEEEtran}
\usepackage{paralist}
\usepackage{pgf}
\usepackage{epstopdf}
\usepackage{etex}
\usepackage{tikz}
\usetikzlibrary{arrows,automata}
\usepackage{algcompatible}
\usepackage{algorithm}
\usepackage{soul}
\usepackage{amsmath}

\usepackage{amssymb}
\usepackage{float}
\usepackage{array}
\usepackage[american]{babel}
\usepackage{graphicx}
\usepackage{subfigure}
\usepackage{amssymb}
\usepackage{enumitem}
\usepackage{algorithm}
\usepackage{url}
\usepackage{xcolor,listings}
\usepackage{verbatimbox}
\usepackage{multirow}
\usepackage{tikz}
\usepackage{pgfplots}
\usepackage{tikz-qtree}
\usepackage{sgame} 

\usepackage{amsthm}

\usepackage{multirow,array}
\usepackage{algorithm}
\usepackage[noend]{algpseudocode}
\usepackage{forest}
\usetikzlibrary{matrix,calc,positioning}
\usetikzlibrary{arrows.meta, shapes.geometric, calc, shadows}
\colorlet{mygreen}{green!75!black}
\colorlet{col1in}{red!30}
\colorlet{col1out}{red!40}
\colorlet{col2in}{mygreen!40}
\colorlet{col2out}{mygreen!50}
\colorlet{col3in}{blue!30}
\colorlet{col3out}{blue!40}
\colorlet{col4in}{mygreen!20}
\colorlet{col4out}{mygreen!30}
\colorlet{col5in}{blue!10}
\colorlet{col5out}{blue!20}
\colorlet{col6in}{blue!20}
\colorlet{col6out}{blue!30}
\colorlet{col7out}{orange}
\colorlet{col7in}{orange!50}
\colorlet{col8out}{orange!40}
\colorlet{col8in}{orange!20}
\colorlet{linecol}{blue!60}
\usepackage{listings}

\colorlet{punct}{red!60!black}
\definecolor{background}{HTML}{EEEEEE}
\definecolor{delim}{RGB}{20,105,176}
\colorlet{numb}{magenta!60!black}

\lstdefinelanguage{json}{
    basicstyle=\normalfont\ttfamily,
    numbers=left,
    numberstyle=\scriptsize,
    stepnumber=1,
    numbersep=8pt,
    showstringspaces=false,
    breaklines=true,
    frame=lines,
    backgroundcolor=\color{background},
    literate=
     *{0}{{{\color{numb}0}}}{1}
      {1}{{{\color{numb}1}}}{1}
      {2}{{{\color{numb}2}}}{1}
      {3}{{{\color{numb}3}}}{1}
      {4}{{{\color{numb}4}}}{1}
      {5}{{{\color{numb}5}}}{1}
      {6}{{{\color{numb}6}}}{1}
      {7}{{{\color{numb}7}}}{1}
      {8}{{{\color{numb}8}}}{1}
      {9}{{{\color{numb}9}}}{1}
      {:}{{{\color{punct}{:}}}}{1}
      {,}{{{\color{punct}{,}}}}{1}
      {\{}{{{\color{delim}{\{}}}}{1}
      {\}}{{{\color{delim}{\}}}}}{1}
      {[}{{{\color{delim}{[}}}}{1}
      {]}{{{\color{delim}{]}}}}{1},
}

\ifCLASSINFOpdf
\else
\fi

\begin{document}
\newtheorem{mydef}{Definition}

%
\title{Adaptive MTD Security using Markov Game Modeling}
\author{
	\IEEEauthorblockN{Ankur Chowdhary, Sailik Sengupta, Adel Alshamrani, Dijiang Huang, and Abdulhakim Sabur}
	\IEEEauthorblockA{Arizona State University
		\\\{achaud16, sailiks, aalsham4, dijiang, asabur\}@asu.edu}
}

\maketitle

\begin{abstract}

Large scale cloud networks consist of distributed networking and computing elements that process critical information and thus security is a key requirement for any environment. Unfortunately, assessing the security state of such networks is a challenging task and the tools used in the past by security experts such as packet filtering, firewall, Intrusion Detection Systems (IDS) etc., provide a reactive security mechanism. In this paper, we introduce a Moving Target Defense (MTD) based proactive security framework for monitoring attacks which lets us identify and reason about multi-stage attacks that target software vulnerabilities present in a cloud network. We formulate the multi-stage attack scenario as a two-player zero-sum Markov Game (between the attacker and the network administrator) on attack graphs. The rewards and transition probabilities are obtained by leveraging the expert knowledge present in the Common Vulnerability Scoring System (CVSS). Our framework identifies an attacker's optimal policy and places countermeasures to ensure that this attack policy is always detected, thus forcing the attacker to use a sub-optimal policy with higher cost.
\end{abstract}

\IEEEpeerreviewmaketitle

\section{Introduction}
   \noindent  A cloud data center consists of software and services from various vendors. Although the security policies of an organization might be up to date, vulnerabilities in software and presence of untrusted insiders can put sensitive information and communication in the network at risk.

    \noindent Traditional defense mechanisms in networks are composed of distributed elements such as firewalls, Intrusion Detection Systems (IDS), log monitoring systems, etc. Also, most of the defense mechanisms are based on reactive/incident-response mechanism. In a modern-era, such an approach can lead to loss of business. Therefore, we need a pro-active approach that anticipates potential weak links in security and assesses the possible behavior of the attacker, in effect providing a defense mechanism that will lead to increased difficulty for an attacker to exploit the network.

    \noindent One such approach that has emerged based on pro-active defense is known as Moving Target Defense (MTD). The goal in network-based MTD is to reconfigure the network services and connectivity in a way that any strategy devised by attacker based on a static view of the network becomes less effective. MTD based adaptive security can increase exploitation surface and decrease attack surface compared to a static system.

    \noindent An ad-hoc approach of switching services and connections in the network can prove to be more catastrophic than useful. Thus, some intelligent scheme is required to take such key decisions. We need to perform some cost-intrusiveness analysis before taking some decision that can have a cascading effect on dependent system components. Game Theory has proved to be very effective in economics, biology, and other areas for taking some important decisions. In this research work, we utilize dynamic game of perfect information between the attacker and the system administrator to create an MTD strategy against attackers targeting software vulnerabilities. 

    The key contribution of this research work are as follows:
    \begin{itemize}
        \item Dynamic Game to perform MTD attack analysis and countermeasure evaluation. Most of the works we came across consider security as a static game. The dynamic game we present mimics realistic network security scenario where attack and defense is a continuous process.
        \item Optimal countermeasure selection, which identifies the critical nodes in an Attack Graph based on CVSS score~\cite{cvss} as a reward metric, and selectively applies countermeasure to mitigate security threats in a cloud network.   
    \end{itemize}

\section{Background}
\label{sec:background}
    \begin{mydef}
        A vulnerability is a security flaw in a software service hosted over a given port, that when exploited by a malicious attacker, can cause loss of Confidentiality, Availability or Integrity (CIA) for a virtual machine (VM).
    \end{mydef}

  \subsection{Threat Model}   
    
    \begin{figure}[!ht]
     \centering
      \includegraphics[width=0.50\textwidth]{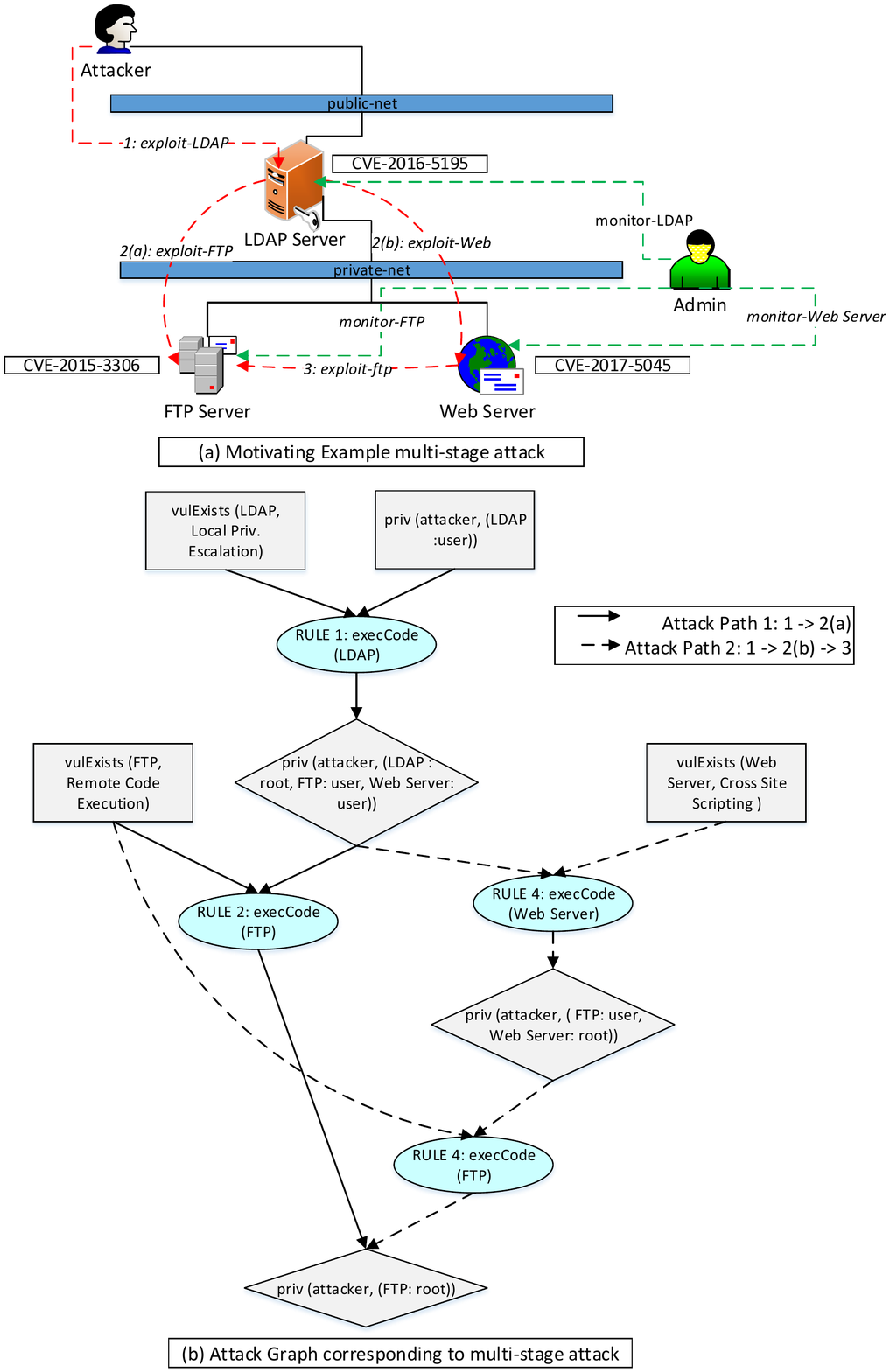}
      \vspace{-2em}
      \caption{An example cloud network scenario.}
      \label{fig:scenario1}
      \vspace{-0.8em}
    \end{figure}
     
     \noindent Consider the cloud system in Figure~\ref{fig:scenario1}(a), where the attacker has user level access to the LDAP server, which is the initial state of our game and the goal state is to compromise the FTP server. The attacker can perform actions such as \textit{1: exploit-LDAP}, \textit{exploit-Web}, \textit{exploit-FTP}. In the scenario above the attacker has two possible paths to reach the goal node \textit{priv(attacker, (FTP: root))}, i.e.
     
     \begin{itemize}
     \item Path 1: exploit-LDAP $\rightarrow$ exploit-FTP 
     \item Path 2: exploit-LDAP $\rightarrow$ exploit-Web $\rightarrow$ exploit-FTP
     \end{itemize}
     
     \noindent The Admin, on the other hand, can choose to monitor (1) running services, (2) network traffic along both the paths using network and host-based monitoring agents, e.g., \textit{monitor-LDAP}, \textit{monitor-FTP}, etc. We assume that the Admin has a limited budget and thus must try to perform monitoring in an optimized fashion. On the other hand, the attacker should try to perform attacks along the path not monitored by the Admin. For example, if the attacker is monitoring traffic only between LDAP and FTP server (mon-LDAP, mon-FTP), the attacker can choose \textit{Path 2} to avoid detection. 
         
     \noindent Attack Graphs (AGs) have proved to be a successful tool for modeling attack behavior. Sheyner et al.~\cite{sheyner2002automated} have discussed a framework for analyzing attacks using formal methods, in which they model the attacker's behavior as an MDP. Unfortunately, the authors have not studied the impact of deploying MTD countermeasures on normal services of the system. We utilize attack graphs to define actions of Attacker and Admin over different stages of the network.\\   
    \textbf{Attack Graph} $G=\{N,E\}$ consists of a set of nodes (N) and a set of edges (E) where,
    \begin{itemize}
    \item As shown in the Figure~\ref{fig:scenario1}(b), the nodes (N) of attack graph can be denoted by $ N = \{ N_f \cup N_c \cup N_d \cup N_r \} $. Here $N_f$ denotes primitive/fact nodes e.g. vulExists (LDAP, Local Priv. Escalation), $N_c$ denotes the exploit, e.g., execCode(LDAP), $N_d$ denotes the privilege level, e.g., priv(attacker, (LDAP :user)) and $N_r$ represents the root or goal node, e.g., priv(attacker, (FTP: root));
    \item The edges (E) of the attack graph can be denoted by $ E =  \{E_{pre} \cup E_{post} \}$. Here $E_{pre} \subseteq (N_f \cup N_c)  \times (N_d \cup N_r)$ ensures that pre-conditions $N_c$ and $N_f$ must be met to achieve $N_d$ and $E_{post} \subseteq (N_d \cup N_r) \times (N_f \cup N_c)$ means post-condition $N_d$ achieved on satisfaction of $N_f$ and $N_c$.
    \end{itemize}
      
    \begin{table}[htb]
    \centering
    \caption{Vulnerability Information for the Cloud Network}
    \label{tab:2}
    \begin{tabular}{ p{1cm} | p{2cm} | p{1.4cm} | p{0.8cm} | p{1.3cm}}
    \hline
    \textbf{VM} & \textbf{Vulnerability} & \textbf{CVE} & \textbf{CIA} & \textbf{AC}  \\ 
    \hline
    LDAP & Local Priv Esc & CVE-2016-5195 & 5.0 & MEDIUM\\ 
    \hline
    Web Server & Cross Site Scripting & CVE-2017-5095 & 7.0 & EASY \\
    \hline
    FTP & Remote Code Execution & CVE-2015-3306 & 10.0 & MEDIUM\\ 
    \hline 
    \end{tabular} 
    \end{table}
    
    \noindent The Table~\ref{tab:2}, shows the Access Complexity (AC) which represents how difficult it is to exploit a vulnerability, and,  Confidentiality, Integrity, and Availability (CIA) gained by exploiting the vulnerabilities in the cloud network above. The values of AC are categorical \texttt{\{EASY, MEDIUM, HIGH\}}, while CIA values are in the range $[0, 10]$.

 \subsection{Game Theoretic Modeling}
        
\noindent \textbf{Players: }We consider the cloud system to be a dynamic game with imperfect information between two players (Attacker:$P_1$ and Admin:$P_2$). $P_1$ is located outside the cloud network, or is a stealthy attacker having user-level access on a particular VM in the cloud network. $P_2$ has the global view of the network. 

\noindent \textbf{Goals: }The goal of the attacker is also well defined, which in this case is to obtain \textit{root} privileges on critical resource(s) of the cloud network like the FTP-Server. We consider the system architecture and a particular use case as shown in Figure~\ref{fig:scenario1}(a). Further, the attack mounted by the attacker is considered monotonic, i.e., once an attacker has reached a certain state, they do not need to go back to any previous state, when targeting a specific goal.

\noindent \textbf{States: }represent the privilege attacker/defender currently have in the network over different resources. We extract the information from the network attack graph to define the state information, e.g., for the attacker, initial state $s_1 = (LDAP, user)$, on the successful execution of the \textit{exploit-LDAP}, the attacker can transition to another state $s_2 = (LDAP, root)$. 

\noindent \textbf{Actions and State Transitions: }The $P_1$ has two possible actions in each state for the example defined in the Figure~\ref{fig:scenario1}, e.g., $a_1^1$ = \textit{no-action}, $a_1^2$ = \textit{exploit-LDAP}. Similarly the $P_2$ has possible actions to monitor LDAP, i.e., $a_2^1$ = \textit{mon-LDAP},  or perform no monitoring, i.e., $a_2^2$ = \textit{no-mon}. 

\begin{figure}[!ht]
     \centering
      \includegraphics[width=0.45\textwidth]{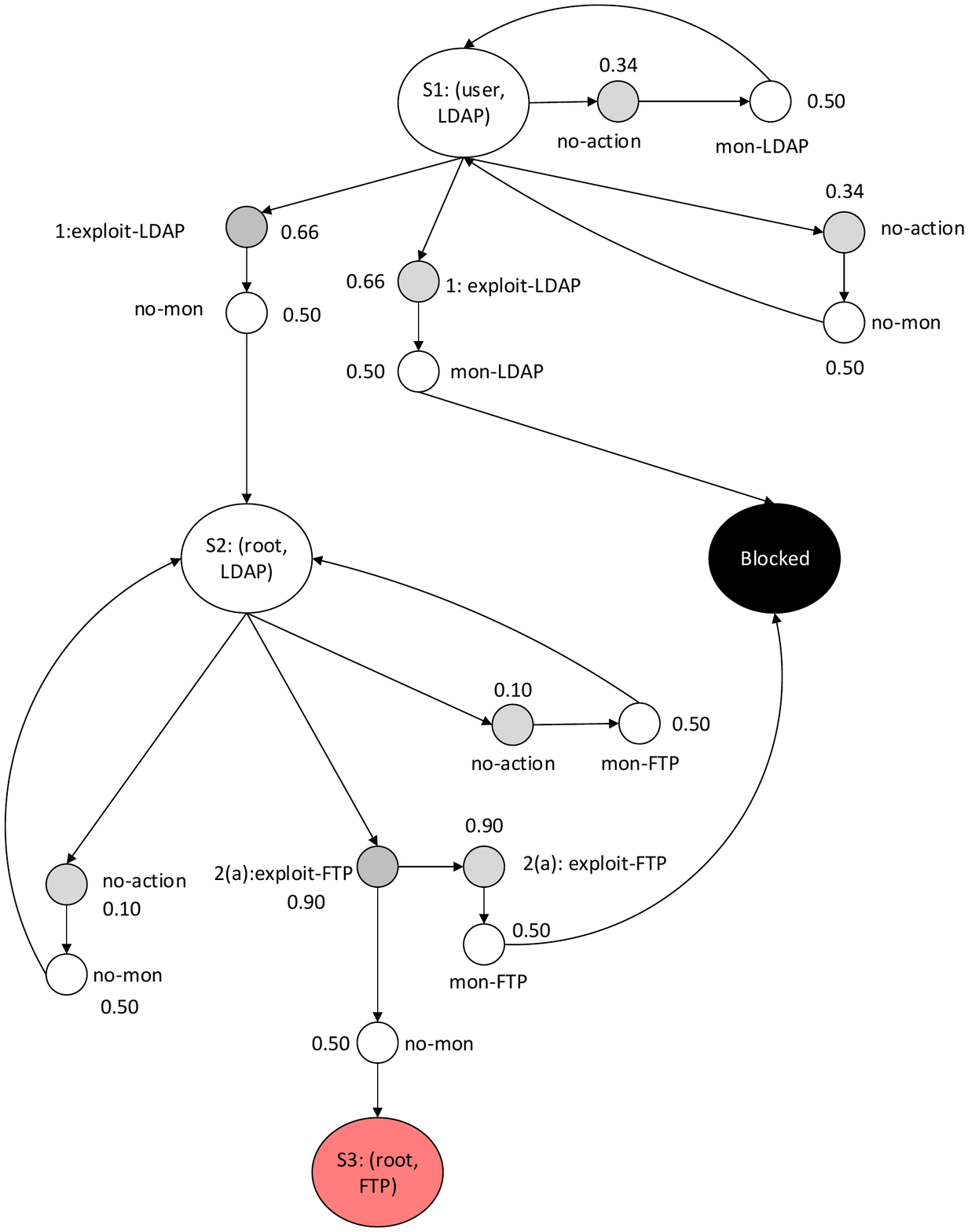}
      \vspace{-3.0em}
     \caption{Actions and State Transitions for Markov Game along Path 1}
     \vspace{-1.0em}
      \label{fig:trans}
\end{figure}

\noindent The Figure~\ref{fig:trans}, shows the probabilities of players $P_1$, $P_2$ actions in each state of the attack along the Path 1, exploit-LDAP $\rightarrow$ exploit-FTP. The state-transition in the Markov Game is conditioned upon the actions of both players, in each state, as shown above. Initially the attacker is present in the state (user, LDAP). The attacker can choose to take an action from the set $\{$no-exploit, exploit-LDAP$\}$ and the probability of taking action exploit-LDAP is 0.66 for the attacker. Similarly, the Admin, has two possible actions, i.e., $\{$no-mon, mon-LDAP$\}$. The admin performs mon-LDAP with a probability 0.5. 

\noindent If the attacker is in state 's', then $\tau(s, a_1, a_2, s')$ is the next state of the game provided $P_1$ and $P_2$ take actions $a_1$ and $a_2$ in the state $s$. In the example above the attacker is only able to exploit LDAP vulnerability if $\tau(s_1$=(LDAP, user), $a_1$=exploit-LDAP, $a_2$=no-mon, $s_2$=(LDAP,root)) $> 0$.

\textbf{Rewards: }The reward function is dependent upon the actions of the attacker and the defender in each state. We refer to the CIA values defined in Table~\ref{tab:2} for the reward associated with successful/unsuccessful action of $P_1$ and $P_2$.

\begin{table}[h]
	\footnotesize
	\begin{tabular}{cc|c|c|c|}
		& \multicolumn{1}{c}{} & \multicolumn{3}{c}{$P_2$}\\
		& \multicolumn{1}{c}{} & \multicolumn{1}{c}{no-mon}  & \multicolumn{1}{c}{mon-Web} & \multicolumn{1}{c}{mon-FTP} \\\cline{3-5}
		& no-act & $0,0$ & $0.5,-0.5$ & $0.5,-0.5$ \\\cline{3-5}
		\multirow{1}*{$P_1$} & exp-Web & $7,-7$ & $-7,7$ & $7,-7$ \\\cline{3-5}
		& exp-FTP & $10,-10$ & $10,-10$ & $-10,10$ \\\cline{3-5}
	\end{tabular}
	\vspace{4pt}
	\caption{Table for state $s_2$}
\end{table}
\vspace{-26pt}
\begin{table}[h]
	\footnotesize
	\begin{tabular}{c|c|c|}
		\multicolumn{1}{c}{} & \multicolumn{1}{c}{no-mon}  & \multicolumn{1}{c}{mon-LDAP} \\\cline{2-3}
		no-act & $0,0$ & $0.5,-0.5$ \\\cline{2-3}
		LDAP & $5,-5$ & $-5,5$ \\\cline{2-3}
	\end{tabular}
	\begin{tabular}{c|c|c|}
		\multicolumn{1}{c}{} & \multicolumn{1}{c}{no-mon}  & \multicolumn{1}{c}{mon-LDAP} \\\cline{2-3}
		no-act & $0,0$ & $0.5,-0.5$ \\\cline{2-3}
		LDAP & $10,-10$ & $-10,10$ \\\cline{2-3}
	\end{tabular}
	\vspace{4pt}
	\caption{Table for state $s_0$ (left) and $s_1$ (right).}
	\label{fig:3}
\end{table}
\vspace{-20pt}


\noindent Table~\ref{fig:3}, shows the possible actions in three important states of the attack propagation, and the corresponding reward metric. When the attacker is present in the state (user, LDAP) - Table~\ref{fig:3}(a), the $P_1$ can take actions $\{$no-act, exploit-LDAP$\}$, similarly the $P_2$ can choose to monitor or not monitor the LDAP server, so the actions for $P_2$ are $\{$no-mon, mon-LDAP$\}$. The reward function is defined as R$(s,a_1,a_2,s')$, where s is the current state and $a_1, a_2$ are actions of both players. For instance if $s_1$=(LDAP, user), $a_1$=exp-LDAP, $a_2$=no-mon, $R(s,a_1,a_2,s')$ = 5.0 for the attacker and -5.0 for the Admin. Similarly the reward metrics for other possible states has been defined in the transition tables above.

    \section{Markov Game}
        \noindent We model the scenario between an attacker and an administrator as a two-player zero-sum Markov game leveraging the knowledge in the the Attack Graph (AG) defined in Section~\ref{sec:background}.
        Besides the obvious Markovian assumption, we further assume that this model has (1) states and actions that are both discrete and finite, and (2) transition from each state and the reward in each state depends on the action each player decides to take in that state. We now formally define a zero-sum Markov Game model and then clearly highlight how each of these are obtained in our setting.
		\begin{mydef}
			A Markov game for two players $P_1$ and $P_2$ can be defined by the tuple $(S, A_1, A_2, \tau, R, \delta)$ where,
			\begin{itemize}[nosep]
				\item $S = \{s_1, s_2, s_3, \ldots, s_k\}$ are finite states of the game,
				\item $A_1 = \{a_1^1, a_1^2, \ldots, a_1^m\}$  represents the possible finite action sets for $P_1$,
				\item $A_2 = \{a_2^1, a_2^2, \ldots, a_2^n\}$ are finite action sets for $P_2$,
				\item $\tau(s, a_1, a_2, s')$ is the probability of reaching a state $s' \in S$ for state $s$ if  $P_1$ and $P_2$ take actions $a_1$ and $a_2$ respectively,
				\item  $R (s, a_1, a_2)$ is the reward obtained by $P_1$ if in state $s$, $P_1$ and $P_2$ take the actions $a_1$ and $a_2$ respectively. Note that the reward for $P_2$ is $-R (s, a_1, a_2)$ by definition of a zero-sum game, and
				\item $\gamma \mapsto (0,1]$ is discount factor for future discount rewards.
			\end{itemize}
		\end{mydef}%
        We model each VM in the AG as a state in our game. The attacker is trying to take actions that help it to reach a particular VM, which is the terminal state, while the administrator's action represents placing a monitoring system to detect an attack. Considering placing a VM incurs cost (negative reward) and there might be multiple vulnerabilities in a single VM, both the attacker and defender have multiple actions depending on the state they are in.
        If the attacker tries to exploit a vulnerability in a state for which the defender choose to place a detection system, they will are detected and blocked with high probability (as defined by $\tau$) and gets a negative reward (which implies the defender gets a positive reward). Otherwise, it has higher probability of succeeding in the attack, thus moving closer to the goal state in the AG and obtaining a positive reward.
        A normal-form zero-sum reward matrix for three states in our simple Markov game is shown in Table~\ref{fig:3}.  Note that, in this formulation which is a first step, we assume that the states are visible to both the players, which implies that the defender gets to know if an attacker has succeeded and hence moved on the a new state $s'$ from $s$. We plan to consider partial observability about the state as future work.
        
        In this game, $P_1$ will try to maximize his expected discounted reward, while $P_2$ will try to select actions that minimize the expected reward for $P_1$. We consider the $\min-\max$ strategy for calculating the expected reward of $P_1$ in our Markov game.
        
        Due to the underlying stochasticity of the game, the players have to reason about the expected reward that they will get for making a particular decision in a certain state. Going forward, we will reason about the rewards for $P_1$, i.e. the attacker and by the property of a zero sum game, we will show that an optimal policy for the defender seeks to reduce this reward value. We now define the quality of an action or the $Q$ value used to represent the expected reward $P_1$ will get for choosing action $a_1 \in A_1$ (while $P_2$ chooses $a_2 \in A_2$),
        \begin{eqnarray}
        	\footnotesize
        	Q(s, a_1, a_2) = R(s, a_1, a_2) + \gamma \sum_{s'} \tau(s, a_1, a_2, s') \cdot V(s')
        	\label{eq:Q}
        \end{eqnarray}%
    	Let us now denote the mixed policy for state $s$ as $\pi(s)$, which is a probability distribution that $P_1$ can now have over the possible actions it can take in state $s$. With that, we can define the value of state $s$ for $P_1$ using the equation,
    	\begin{eqnarray}
    		V(s) = \max_{\pi(s)} \min_{a_2} \sum_{a_1} Q(s, a_1, a_2)\cdot \pi_{a_1}
    		\label{eq:V}
    	\end{eqnarray}%
        Note that the the max-min strategy of attacker (and the defender) can be captured by using a modified version of the classic value iteration algorithm. Using the min-max strategy implies that the defender is trying to minimize the reward of then attacker by placing detection nodes in the various states that are a part of the various attack paths present in the attack graph, while the attacker is trying to use a mixed policy $\pi(s)$ over it possible actions in $A_2$ to maximize their total reward.
        Thus, the Markov Game framework helps the administrator model the attackers policy so that they can take necessary countermeasures (decision in each state for player $P_2$) to minimize the expected utility for the attacker.
   
\section{Implementation and Evaluation}

    \subsection{Implementation}
       \label{sec:impl}
       Note that in order to implement the actual modified value iteration (as described by Eq. \ref{eq:Q} and \ref{eq:V}), one needs to solve a linear program for each state after obtaining the updated $Q$ values. This may become computationally inefficient for large real networks and thus, we make two assumptions, that provide us with an approximate strategy for the players. First, we restrict the attacker to select a pure strategy, i.e. $\pi(s) ~\forall s$ has the value of $1$ for only one action $a\in A_2$ that it can execute in state $s$ and zero for all other actions. Second, we consider that the defender has observability of the attacker's action and thus only use action pairs $(a_1, a_2)$ that respect the min-max condition. With these, we can not use the classic value iteration algorithm where the action set of the MDP $A$ is restricted action pairs mentioned above and the reward and transition function is a subset of the original ones for the Markov Game only defined for those action pairs that respect the min-max strategy.

\begin{algorithm}
\caption{MDP - VALUE ITERATION}\label{euclid}
\begin{algorithmic}[1]

\Procedure{Emit- $V^*(S) \quad $ at $t_k$}{}
\State $V_0^*(s) = 0$ for all s. \COMMENT{\{Initialize value function to start\}}
\State $s \gets s_0$
\State $\tau(s,a,s')$  \COMMENT{\{Transition probability from state s to s'\}}
\State $R(s,a,s')$ \COMMENT{\{Reward obtained by taking action a in state s\}}
\State $\delta \in [0,1]$  \COMMENT{\{Discount factor\}}
\State $i \gets 0$
\State \emph{loop}:
\IF{$i == k$}
\STATE break;
\ENDIF
\State $V^*(s) \gets max_{A} \sum_{s'}\tau(s,a,s') \times [R(s,a,s') + \gamma V^*(s')]$
\State $\pi(s) =  argmax_{a} \sum_{s'}\tau(s,a,s') \times [R(s,a,s') + \gamma V^*(s')]$
\State $i \gets i+1$
\State \textbf{goto} \emph{loop}.

\EndProcedure
\end{algorithmic}
\end{algorithm}


\subsection{Optimal Countermeasure Selection}
\noindent We consider countermeasure deployment using the equilibrium strategy for the markov game, which means for each state of the game, the administrator minimizes the maximum benefit to the attacker. We use an OpenStack based cloud environment to make a small system with three VMs, i.e, \textit{a) Ubuntu 16.04: {\tt 192.168.1.5}} b) \textit{Fedora 23: {\tt 192.168.1.6}} c) \textit{metasploitable: {\tt 192.168.1.7}}. In this environment, we assume that the attacker is initially located on the VM with Ubuntu 16.04 and has user-level privileges on it. The attacker conducts a \textit{nmap} scan to enumerate the vulnerable system and network services that are present on the host and the target VM. The attacker's goal is to obtain `root' level privileges on the target VM. We further utilized $cve-search$ \cite{cve-search} to obtain host and guest vulnerability information. 


\noindent We conducted Markov Game cost-benefit analysis for the two players-- Attacker and Admin -- on the system (with $VM_1, VM_2, VM_3$) defined in our evaluation network. We consider $100$ known vulnerabilities spread across the three VMs, that have not yet been fixed due to resource considerations. The goal state for the attacker is to obtain \textit{root} level privileges on all VM's using the available system and network exploits. We conducted experiments to evaluate the effectiveness of strategic MDP Value Iteration based countermeasure vs naive strategy that selects only top x=$\{10,50\}$\% of the vulnerabilities to apply countermeasure. The naive strategy assumes that the administrator should only select VMs with vulnerabilities having high CIA for patching. The strategic approach assumes that the administrator has conducted attack analysis in advance and knows the high-value targets based on MDP value iteration in the cloud network that may be subjected to attack. 

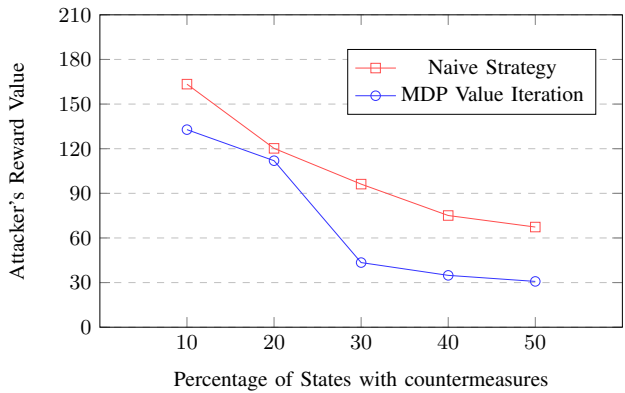
\begin{figure}[t!]
	\centering
	\begin{tikzpicture}[scale=0.9, transform shape]
	\begin{axis}[
	height=0.7\linewidth,
	width=1.05\linewidth,
	font=\small,
	xlabel={Percentage of States with countermeasures},
	ylabel={Attacker's Reward Value},
	xmin=0, xmax=60,
	ymin=0.0, ymax=210,
	xtick={10,20,30,40,50},
	ytick={0, 30.0, 60.0, 90.0, 120.0, 150.0, 180.0, 210.0},
	legend pos=north west,
	ymajorgrids=true,
	grid style=dashed,
	name=border,
	legend style ={
		at={(0.95,0.68)}, 
		anchor=south east,
		draw=black, 
		fill=white,
		font=\small,
	}
	]
	
	\addplot[
	color={red!70},
	mark=square,
	]
	coordinates {
		(10,163.40)(20, 120.23)(30,96.19)(40, 75.09)(50, 67.34)
	};
	
	\addplot[
	color={blue!80},
	mark=o,
	]
	coordinates {
		(10, 132.82)(20, 111.93)(30,43.45)(40,34.86)(50, 30.79)
	};
	\addlegendentry{Naive Strategy};
	\addlegendentry{MDP Value Iteration};
	\end{axis}
	
	\end{tikzpicture}
	\caption{$P_1$ Reward Goal Value vs $P_2$'s  Naive and Strategic countermeasures}
	\label{fig:eval1}
	\vspace{-1em}
\end{figure}

\noindent As shown in the Figure~\ref{fig:eval1} the reward value for Attacker decreases as the administrator increases his attack surface coverage. The reward for attacker decreases from 160 to 96 on providing a countermeasure for 30 percent vulnerable states when a naive strategy is used. The value further decreases from $\sim96$ to $\sim65$ on implementing countermeasure for half of the vulnerable states. The countermeasures are deployed naively based on top vulnerabilities, irrespective of analysis of asset value to an attacker.\\
In the case of a strategic approach, the administrator performs a thorough analysis of vulnerable states, high-value targets, shortest paths to goal states and strategically implements countermeasures for those states. The reward value for attacker decreases from $\sim133$ to 45 for the attacker when administrator increases countermeasures for about 30 percent vulnerable states from 10 percent states. This value further decreases to 30 for the attacker when the administrator has coverage for about 50 percent vulnerable states. \\
We can observe that the strategic approach affects attackers reward significantly, a reduction of almost half. An attacker can obtain the reward of 30 for a strategic approach, compared to reward $\sim65$ for naive approach when administrator deploys countermeasures for 50\% vulnerable states. The interdiction of attack paths by the administrator using Markov Game, being the optimal, will \textit{strictly dominate any} other strategy and thus, significantly limit the attacker's capability as the number of vulnerabilities will increase in the cloud environment.



\vspace{-0.7em}
\section{Related Work}
\label{sec:rel}
Sheyner \textit{et al} \cite{jha2002two} present a formal analysis of attacks on a network along with cost-benefit analysis and security measures to defend against the network attacks. In \cite{chowdhary2016sdn}, Chowdhary {\em et al.} provide a polynomial time method for attack graph construction and network reconfiguration using a parallel computing approach, making it possible to leverage information for strategic reason of attacks in large-scale systems.

Authors in \cite{jia2013motag} introduced the idea of moving secret proxies to new network locations using a greedy algorithm, which they show can thwart brute force and DDoS attacks. In \cite{zhuang2013investigating}, Zhuang {\em et al} shows that MTD system designed with intelligent adaptations improve the effectiveness further. In \cite{sengupta2017game}, authors shows that intelligent strategies based on common intuitions can be detrimental to security and highlight how game theoretic reasoning can alleviate the problem. On those line, Wei \textit{et al} \cite{lye2005game} and Sengupta \textit{et al} \cite{senguptamoving} use a game theoretic approach to model the attacker-defender interaction as a two-player game where they calculate the optimal response for the players using the Nash and the Stackelberg Equilibrium concepts respectively. Although they propose the use of the Markov Decision Process (MDP) and attack graph-based approaches, they leave it as future work.

In the context of cloud systems, Peng {\em et al} discusses a risk-aware MTD strategy \cite{peng2014moving} where they model the attack surface as a non-decreasing probability density function and then estimate the risk of migrating a VM to a replacement node using probabilistic inference. Kampanakis \textit{et al} \cite{kampanakis2014sdn} highlight obfuscation as a possible MTD strategy in order to deal with attacks like OS fingerprinting and network reconnaissance in the SDN environment. Furthermore, they highlight that the trade-off between such random mutations, which may disrupt any active services, require analysis of cost-benefits.

In this paper, we identify an adaptive MTD strategy against multi-hop monotonic attacks for cloud networks which optimizes the performance while providing gains in security.
 \vspace{-0.8em}
\section{Conclusion and Future Work}
\label{sec:concl}
 \noindent A cloud network is composed of heterogeneous network devices and applications interacting with each other. The interaction of these entities poses both (1) a security risk to overall cloud infrastructure and (2) makes it difficult to secure them. While traditional security solutions provide reactive security mechanisms to detect and mitigate a threat, they may fail to assess the damage to infrastructure due to a cascading security breach. We presented Markov Game as an assessment tool to perform a cost-benefit analysis of security vulnerabilities and corresponding countermeasures in the cloud network. 
 The assessment shows that a network administrator needs to proactively identify critical security assets and strategically deploy available countermeasures. Game Theoretic approach will help the administrator to quantify and minimize risk provided limited resources. 
  \vspace{-0.6em}
\section*{Acknowledgment}
\noindent This research is based upon work supported by the NRL N00173-15-G017, NSF Grants 1642031, 1528099, and 1723440, and NSFC Grants 61628201 and 61571375. The second author is supported by the IBM Ph.D. Fellowship.

\bibliographystyle{abbrv}
\bibliography{mtd}

\end{document}